\documentclass[prb,aps,twocolumn]{revtex4}
\usepackage{graphicx}
\usepackage{subfigure}
\usepackage{bm}
\usepackage{amsmath}
\begin{document} 
\title{On dynamical tunneling and classical resonances}
\author{Srihari Keshavamurthy\footnote{Permanent address: Department of
Chemistry, Indian Institute of Technology, Kanpur, U.P. 208016, India.}}
\affiliation{Institut f\"{u}r Theoretische Physik, 
Universit\"{a}t Regensburg,
93040 Regensburg, Germany}
\date{\today}
\begin{abstract}
This letter establishes a firm relationship between classical nonlinear
resonances and the phenomenon of dynamical tunneling. It is shown that
the classical phase space with its hierarchy of resonance islands 
completely characterizes dynamical tunneling. 
In particular, it is not important to invoke criteria
such as the size of the islands
and presence or absence of avoided crossings
for a consistent description of dynamical
tunneling in near-integrable systems.
\end{abstract}
\maketitle

Dynamical tunneling as a concept emerged more than two decades ago
in the field of chemical physics where 
it results in the transport of
vibrational quanta between degenerate modes - a process that
would be classically forbidden.
The importance of dynamical tunneling in the molecular context
can be hardly overstated since this phenomenon provides
a route to energy flow through the molecule in the absence of
direct classically resonant mechanisms.
Early pioneering work\cite{lawch,davhel,husihy,stumar,orti} 
mainly by the chemical physics community provided 
both semiclassical\cite{lawch,davhel}
and purely quantum perspectives\cite{husihy,stumar} on dynamical tunneling.
Semiclassically the phase space is the natural setting whereas
the quantum approach invokes high order perturbation theory
involving a chain of off-resonant virtual states (vibrational
superexchange\cite{stumar}).
Although seemingly different, there are
hints\cite{stumar,self} towards a connection between the two perspectives and
this paper attempts to provide further clues.

The initial suggestion\cite{davhel} regarding the importance
of phase space structures to dynamical tunneling 
has been intensely studied and established by the 
nonlinear dynamics community
over the last decade\cite{ozo,cat,cats,rat1,rat2,frido}.
Dynamical tunneling is found not only to be 
influenced by chaos\cite{cat,cats,frido} 
but also by various nonlinear resonances\cite{rat1,rat2,frido,self}
with some recent experimental support\cite{expt}.
In the molecular context
Heller recently\cite{ejhsar} made
a number of interesting observations and conjectures on the possible
implications of dynamical tunneling on 
high resolution molecular spectra\cite{lehm}.
The most important amongst these is the claim that a nominal 
10$^{-1}$-10$^{-2}$ cm$^{-1}$ broadening of spectroscopically
prepared zeroth order states is due to dynamical tunneling between
remote regions of phase space facilitated by distant resonances.
Arguments were provided for identifying the specific resonances
and subsequent calculation of the splittings. The purpose of this
letter is to confirm the above claim via a detailed analysis of
a relatively simple, albeit realistic, model spectroscopic Hamiltonian.
The analysis also indicates that the correspondence between
classical resonances and avoided crossings, while interesting, is 
not needed for an understanding of dynamical tunneling.

We use a model spectroscopic Hamiltonian\cite{ksjcp} 
\begin{equation}
\hat{H}=\hat{H}_{0}+g \hat{V}^{(12)}_{1:1}+\gamma \hat{V}^{(12)}_{2:2}
+\beta(\hat{V}^{(1b)}_{2:1}+\hat{V}^{(2b)}_{2:1})
\label{eq1}
\end{equation}
with
\begin{eqnarray}
\hat{H}_{0}&=&\omega_{s}(n_{1}+n_{2})+\omega_{b}n_{b}
+x_{s}(n^{2}_{1}+n^{2}_{2}) \nonumber \\
&+&x_{b}n^{2}_{b}+x_{sb}n_{b}(n_{1}+n_{2})+
x_{ss}n_{1}n_{2}
\end{eqnarray}
The values of these parameters, in cm$^{-1}$,  
$\omega_{s}=3885.57,\omega_{b}=1651.72,x_{s}=-81.99,x_{b}=-18.91,
x_{ss}=-12.17$, and $x_{sb}=-19.12$ are representative of the
H$_{2}$O molecule\cite{bag}.
The
three anharmonic modes are labeled as stretches $(1,2)$ and a bend $(b)$
and
$\hat{H}$ is symmetric under $1 \leftrightarrow 2$.
The $j^{th}$ mode occupancy is $n_{j}=a^{\dagger}_{j}a_{j}$
with $(a^{\dagger}_{j},a_{j})$ denoting the usual harmonic oscillator
creation and destruction operators for mode $j$.
The various perturbations 
$\hat{V}^{(ij)}_{p:q}=(a^{\dagger}_{i})^{q}(a_{j})^{p}+h.c.$ 
connect zeroth-order states
$|{\bf n}\rangle$, $|{\bf n}'\rangle$ with $|n_{i}'-n_{i}|=q$ and
$|n_{j}'-n_{j}|=p$. The classical limit\cite{ksjcp} 
of the above Hamiltonian is
a nonlinear multiresonant
Hamiltonian $H({\bf I},{\bm \theta})$
with $(I_{j},\theta_{j})$ corresponding to the action-angle variables
associated with the mode $j$. 
In particular the classical limit of
$\hat{V}^{(ij)}_{p:q}$ is of the form
$2 \sqrt{I_{i}^{q}I_{j}^{p}} \cos(q\theta_{i}-p\theta_{j})$.
$\hat{H}$ can be obtained by a fit to the 
high resolution experimental spectra or from a perturbative analysis of
a high quality {\it ab initio} potential energy surface. In either case
such effective Hamiltonians provide a very natural
and convenient
representation to understand the spectral patterns of molecular
systems\cite{hrs}. 
Note that despite the three coupled modes, $\hat{H}$
is effectively two dimensional due to the existence of the
conserved quantity $P=(n_{1}+n_{2})+n_{b}/2$ called as the polyad number.
The classical Hamiltonian is integrable if $\beta=0$ and
for our choice of parameters it is near-integrable if $\beta \neq 0$.
Throughout this study we fix $P=8$ and $\beta=26.57$ cm$^{-1}$ 
and denote, due to conserved $P$,
the zeroth-order states by $|n_{1},n_{2}\rangle$. 

\begin{figure}
\includegraphics*[width=3in,height=3in]{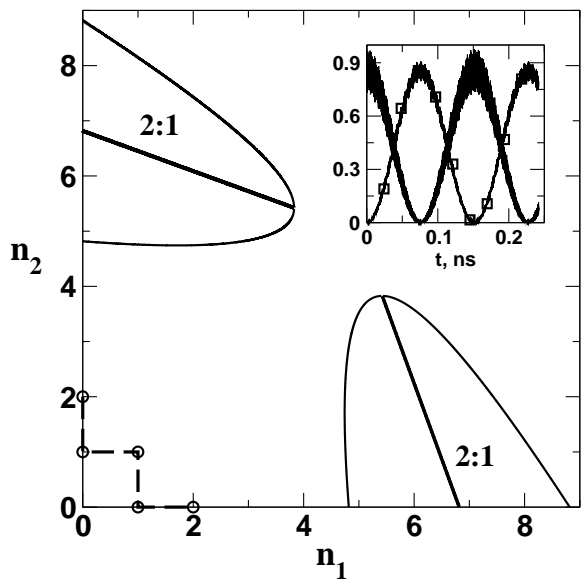}
\caption{State or zeroth-order number space
with the location and width
of the 2:1 resonance zones. An example
superexchange path (dashed) between the degenerate states
$|2,0\rangle$
and $|0,2\rangle$ is indicated. The inset shows the survival
probability for $|2,0\rangle$ and $|0,2\rangle$ (symbols)
versus time.}
\label{fig1}
\end{figure}

To begin with consider the case wherein only the 2:1 resonances are
present {\it i.e.,} $g=\gamma=0$. 
In order to emphasize and illustrate the concept we
choose the zeroth-order degenerate states $|2,0\rangle$ and $|0,2\rangle$
without loss of generality.
Since $\langle 2,0|\hat{V}^{(1b)}_{2:1}|0,2\rangle = 0 =
\langle 2,0|\hat{V}^{(2b)}_{2:1}|0,2\rangle$ 
the state $|2,0\rangle$ is uncoupled from the
symmetric counterpart $|0,2\rangle$. However, dynamical tunneling
can mix these states 
and indeed from Fig.~\ref{fig1} one observes a coherent transfer of
population with a period of about $0.15$ ns corresponding to
to a splitting $\Delta_{2} \approx 0.22$ cm$^{-1}$.
The nontrivial nature of this process is amplified when one
considers the fact
that at the energy corresponding to $|2,0\rangle$ the
primary 2:1 resonances are absent in the classical phase space.
As shown in Fig.~\ref{fig1} the two states are far away in state space
from the 2:1 primary resonance zones. One possible explanation
of the tunneling arises from the perspective of high order perturbation
theory or vibrational superexchange\cite{husihy,stumar}. 
In this approach the
states coupled locally by the 2:1 perturbations are considered
and one constructs perturbative chains which connect the two
states $|2,0\rangle$ and $|0,2\rangle$. An example of such
a chain is $|2,0\rangle \rightarrow |1,0\rangle \rightarrow
|1,1\rangle \rightarrow |0,1\rangle \rightarrow |0,2\rangle$.
The contribution to the splitting from the chain is given
by perturbation theory to be 
\begin{equation}
\beta^{4}\frac{\langle 20|\hat{V}^{(1b)}|10\rangle \langle
10|\hat{V}^{(2b)}|11\rangle \langle 11|\hat{V}^{(1b)}|
01\rangle \langle 01|\hat{V}^{(2b)}|02\rangle}{
(\Delta E^{0}_{1,0,14})^{2}(\Delta E^{0}_{1,1,12})}
\end{equation}
with $\Delta E^{0}_{n_{1},n_{2},n_{b}} \equiv
E^{0}_{2,0,12}-E^{0}_{n_{1},n_{2},n_{b}}$.
In principle there are an infinite number of chains that connect the
two degenerate states. In practice, due to the energy denominators and
near-integrability, it
is sufficient to consider the 
minimal length chains\cite{self}. In our case
there are six minimal chains and summing
the contributions from each one of them one obtains a splitting of
about $0.23$ cm$^{-1}$. This compares well with the exact splitting
but it is important to note that all six perturbative terms have to be
considered for this agreement. 
Note that although formally the superexchange approach invokes the
resonant terms the connection to the classical phase space
is lost. If indeed dynamical tunneling is properly understood in the
phase space then surely there must be a phase space analog of the
superexchange approach. The rest of the paper is dedicated to
uncovering precisely such a phase space picture.

\begin{figure}
\includegraphics*[width=3in,height=3in]{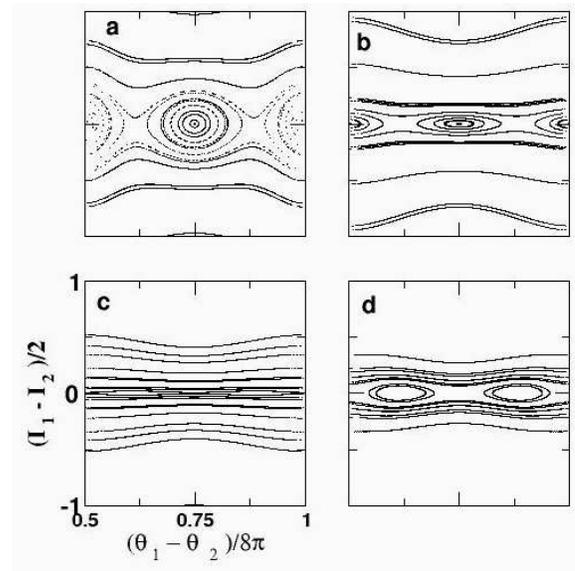}
\caption{Surface of section at $E=E^{0}_{2,0,12}$ and $\beta=26.57$
cm$^{-1}$ for varying primary 1:1 strength in cm$^{-1}$.
(a) $g=0$, (b) $g=-2.7$,
(c) $g=-2.95 \approx g_{0}$, and (d) $g=-3.4$. Note that for $g < g_{0}$
the island corresponds to the induced 1:1 whereas for $g > g_{0}$
the island correspond to the primary 1:1. Also note that a Husimi
representation of the states $|2,0\rangle$ and $|0,2\rangle$ would
be localized about $(I_{1}-I_{2})/2=\pm 1$ respectively.}
\label{fig2}
\end{figure}

As mentioned the primary 2:1 resonances do not appear
in the phase space at 
$E=E^{0}_{2,0,12}$ 
and hence a direct involvement is ruled out. Nevertheless
a weak overlap between the 2:1s can result
in an induced 1:1 resonance which can then
mediate dynamical tunneling between the states. In Fig.~\ref{fig2}a
we show the surface of section at $E=E^{0}_{2,0,12}$ and one indeed observes
a resonance island between the two states. 
In order to confirm the nature of this resonance zone we use 
standard methods of nonlinear dynamics\cite{licht} 
to extract the necessary information.
In essence one starts with the classical Hamiltonian involving only the
2:1 perturbations in the form\cite{ksjcp}
\begin{eqnarray}
H({\bf J},{\bm \psi};N)&=&H_{0}({\bf J};N) +\epsilon
\beta_{c}(N-2J_{1}-2J_{2}) \\
&\times& [\sqrt{J_{1}}\cos\psi_{1}+\sqrt{J_{2}}\cos\psi_{2}] \nonumber
\end{eqnarray}
with $\beta_{c}=\beta/\sqrt{2}$ and 
$N$ being the classical analog of the polyad number. 
A formal parameter $\epsilon$ has been introduced with the
aim of perturbatively removing the 2:1 resonances, characterized by
$\psi_{1,2}$, to $O(\epsilon)$. This can be done by invoking
the generating function 
$F=\bar{J}_{1}\psi_{1}+
\bar{J}_{2}\psi_{2}+\epsilon[g_{1}\sin\psi_{1}+g_{2}\sin\psi_{2}]$
where the functions $g_{1,2}=g_{1,2}(\bar{J}_{1},\bar{J}_{2})$
are determined by the condition of the removal of the primary
2:1s to $O(\epsilon)$. 
The angles conjugate to ${\bar{\bf J}}$ are denoted by
${\bar{\bm \psi}}$.
The procedure is algebraically tedious and we
skip the details to provide the important results. The choice of the
functions $g_{1,2}$ turns out to be:
\begin{equation}
g_{1,2}(\bar{J}_{1},\bar{J}_{2})=-\beta_{c} \frac{(N-2\bar{J}_{1}-2\bar{J}_{2})
\sqrt{\bar{J}_{1,2}}}{\Omega_{s}+2\alpha_{ss}\bar{J}_{1,2}+
\alpha_{12}\bar{J}_{2,1}}
\end{equation}
where $\Omega_{s}=\omega_{s}-2\omega_{b}+(x_{sb}-4x_{b})N,
\alpha_{ss}=x_{s}+4x_{b}-2x_{sb}$ and 
$\alpha_{12}=8x_{b}-4x_{sb}+x_{ss}$. 
Using the above result it is possible to show that an induced 1:1 
resonance appears at $O(\epsilon^{2})$ with a coefficient
\begin{equation}
g_{ind}=\frac{\beta_{c}^{2}}{2}(N-2\bar{J}_{1}-2\bar{J}_{2})
f(\bar{J}_{1},\bar{J}_{2};N) \sqrt{\bar{J}_{1}\bar{J}_{2}}
\end{equation}
with $f(\bar{J}_{1},\bar{J}_{2};N)$ being a complicated function
of the actions. At this stage the transformed Hamiltonian to $O(\epsilon^{2})$
still depends on both the angles $\bar{\psi}_{1}$ and $\bar{\psi}_{2}$ and
hence non-integrable. In order to isolate the induced 1:1 resonance 
we perform a canonical transformation to the variables $({\bf K},{\bm \phi})$
using the generating function $G=(\bar{\psi}_{1}-\bar{\psi}_{2})K_{1}/2 +
(\bar{\psi}_{1}+\bar{\psi}_{2})K_{2}/2$ and average the resulting
Hamiltonian over the fast angle $\phi_{2}$. The resonance center,
$K_{1}^{r}=0$,
approximation is invoked resulting
in a pendulum Hamiltonian describing
the induced 1:1 resonance island structure seen in the surface
of section shown in Fig.~\ref{fig2}a. Within the averaged approximation the
action $K_{2}=\bar{J}_{1}+\bar{J}_{2}$ is a constant of the motion
and can be identified as the 1:1 polyad associated with the secondary
resonance. 
The resulting integrable Hamiltonian is given by
\begin{equation}
\bar{H}(K_{1},\phi_{1};K_{2},N)=\frac{1}{2 M_{11}} K_{1}^{2}+
2g_{ind}(K_{2},N)\cos2\phi_{1}
\end{equation}
where
\begin{subequations}
\begin{eqnarray}
M_{11}&=&2(\alpha_{12}-2\alpha_{ss})^{-1} \\
g_{ind}(K_{2},N)&=&\frac{\beta_{c}^{2}}{2}\bar{f}(K_{2},N)
(N-2K_{2}) K_{2}
\end{eqnarray}
with
\begin{equation}
\bar{f}(K_{2},N)=\frac{4(\Omega_{s}+\alpha_{ss}K_{2})+\alpha_{12}N}
{[2(\Omega_{s}+\alpha_{ss}K_{2})+\alpha_{12}K_{2}]^{2}}
\end{equation}
\end{subequations}
In terms of the zeroth-order quantum numbers $K_{1}=n_{1}-n_{2}$ and
$K_{2}=n_{1}+n_{2}+1 \equiv m+1$.

One can now use the above pendulum Hamiltonian to calculate the
resulting dynamical tunnel splitting 
of the degenerate modes $|n_{1}=r,n_{2}=0,n_{b}=2(P-r)\rangle$
and $|n_{1}=0,n_{2}=r,n_{b}=2(P-r)\rangle$ via\cite{stumar}
\begin{equation}
\frac{\Delta^{sc}_{r}}{2}
=g_{ind} \prod_{m=-(r-2)}^{(r-2)} \frac{g_{ind}}{E_{R}^{0}(r)-
E_{R}^{0}(m)}
\end{equation}
where $E_{R}^{0}(k)=k^{2}/2M_{11}$ is the zeroth-order energy.
For our example with $r=2,m=2$ using the parameters of the Hamiltonian
we find $M_{11} \approx 1.32 \times 10^{-2}$ and $g_{ind} \approx 4.43$
cm$^{-1}$. The resulting splitting $\Delta_{2}^{sc} \approx 0.26$
cm$^{-1}$ agrees very well with the exact splitting.
This proves that the induced 1:1 resonance arising from the
interaction of the two primary 2:1 resonances is mediating dynamical
tunneling between the degenerate states. 
At this juncture it is important to note that the induced resonance
strength is quite small and the two states are not involved in
any avoided crossing. Moreover, from a superexchange perspective
it is illuminating to note that the splitting can be calculated
trivially by recognizing the secondary phase space structure (a viewpoint
emphasized in Ref.~\onlinecite{rat2} as well). 
In
comparison
the original superexchange calculation, without
any reference to the phase space, required taking into
account $6$ terms with varying signs\cite{comment}. 
This observation emphasizes
the superior nature of a phase space viewpoint on dynamical tunneling.

\begin{figure}
\includegraphics*[width=3in,height=3in]{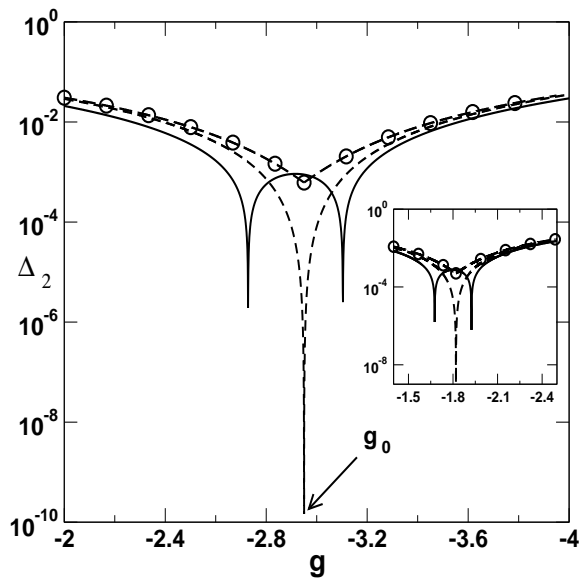}
\caption{The variation of the dynamical splitting $\Delta_{2}$ with
the primary 1:1 resonance strength $g$ is shown (solid line).
The WKB estimate is shown as dashed line. WKB estimate with
a small 2:2 resonance added is also shown (circles). The inset shows
a similar calculation with a different set of parameters
representing the D$_{2}$O molecule\cite{bag}. In this case
$\beta=15.78$ and, $g_{ind}\approx 2.73$ cm$^{-1}$.}
\label{fig3}
\end{figure} 

As a further demonstration of the role of nonlinear resonances
in dynamical tunneling 
we consider the Hamiltonian in eq.~\ref{eq1} 
with $g \neq 0$ with $\gamma = 0$. In particular the sign of the
primary 1:1 perturbation strength $g$ is taken to be the opposite of
the induced 1:1 strength $g_{ind}$. If the induced resonance
is playing a role then from our analysis we expect that the
primary and induced resonances will cancel each other around
$g = g_{0} \approx -2 g_{ind}/(m+1)$ resulting
in small splittings in this region. In Fig.~\ref{fig3}
the exact splittings are shown as a function of $g$ with 
the WKB results\cite{husihy,stumar}
for comparison. This confirms our expectations to a certain
degree in that the splittings are undergoing dramatic changes in the vicinity
of $g_{0}$. A crucial observation is that the exact splitting
is orders of magnitude larger than the simple WKB estimate
and become small slightly away from $g_{0}$. 
On the other hand 
the classical phase space in Fig.~\ref{fig2} indicates the predicted
disappearance of the 1:1 islands. 
From our arguments this far it would be natural to associate one or more
high order nonlinear resonances with the residual tunneling around
$g_{0}$ since the simple semiclassical estimate
for $g_{0}$ was based on the $O(\epsilon^{2})$ induced 1:1 resonance
cancelling the primary 1:1 resonance. In reality there are the 
harmonics of the 1:1 resonance that appear at higher orders in
$\epsilon$. It is expected that the strengths of
such higher harmonics like 2:2, 3:3, etc. would be extremely small.
Nevertheless around $g = g_{0}$ the most dominant resonance involved in
dynamical tunneling would be the 2:2. 
The strength of this tiny but dominant 2:2 
resonance can be estimated roughly by adding a 2:2
perturbation ($\gamma$) to eq.~\ref{eq1} with $\beta=26.57$ cm$^{-1}$,
$g=g_{0}$ and noting the value of $\gamma$ for which the exact and
WKB results come close. A much more rigorous estimate, which
is a difficult excersise in classical perturbation theory, can be
made by going to higher orders, atleast $O(\epsilon^{3})$, in $\epsilon$.
We now estimate the splitting with a WKB calculation 
including the 2:2 resonance with strength $\gamma \approx 2.32355 \times
10^{-4}$ cm$^{-1}$. It is clear from Fig.~\ref{fig3} that the exact splitting
and the modified WKB estimate based on the higher order 2:2 agree
fairly well. It is also satisfying to see that the modified WKB
calculation hardly effects the splittings far away from $g = g_{0}$.
As an independent check in Fig.~\ref{fig3}(inset) we show
the same calculation for a different set of $\hat{H}_{0}$ parameters
representative of the D$_{2}$O molecule\cite{bag} and the results
are similar.
This supports the argument that in the vicinity of $g_{0}$, where the
1:1 resonance is absent, the extremely small 2:2 resonance is mediating
the dynamical tunneling between the states $|2,0\rangle$ and
$|0,2\rangle$. Two remarks are in order at this stage. First the two
states do not undergo any avoided crossing as a function of the
parameter $g$. 
This can also be indirectly inferred from the fact that a superexchange
calculation of the splittings 
essentially reproduces the exact result
and thus include the importance of
higher order resonances near $g=g_{0}$. A more detailed analysis
of the perturbative chains from the semiclassical
viewpoint would be interesting.
Second the modified WKB calculation is in good agreement
with the exact splittings only in the vicinity of $g_{0}$ by necessity.
There are contributions from even higher order resonances which are
absent from our simplified analysis and a subtle interplay of all the
nonlinear resonances give rise to the exact result. 

To summarize, in this work using a model spectroscopic Hamiltonian we
have demonstrated the intimate connection between dynamical tunneling
and the resonance structure of the classical phase space. Thus dynamical
tunneling connects two degenerate states as
long as there is a nonlinear resonance juxtaposed between them
as viewed in the phase space. The
order and width of the resonance are immaterial. This supports an
earlier claim regarding the possibility of dynamical tunneling as a
source of narrow spectral clusters associated with spectroscopically
prepared, localized, zero-order states.
However the notion that such resonances are the cause of avoided
crossings does not seem to hold. Consequently it is also not necessary that 
only a specific classical resonance be the agent of dynamical tunneling.
Primary, induced and even higher harmonics of the resonances can
mediate dynamical tunneling and the consequences for energy flow and
control from this standpoint seem crucial and needs further study.
It is interesting to note that in multidimensional 
near-integrable systems nonlinear
resonances would be involved in two long time phenomena - dynamical
tunneling and Arnol'd diffusion\cite{licht}. 
The competition between them and their
spectral consequences are worth investigating from a fundamental 
standpoint\cite{compet}.

It is a pleasure to thank Peter Schlagheck for critical 
and illuminating discussions.
I am grateful to Prof. Klaus Richter 
for the hospitality and support at the Universit\"{a}t Regensburg where
this work was done.


\begin{thebibliography}{}
\bibitem{comment}{This observation is quite dramatic whe one considers
the dynamical tunneling between $|4,0\rangle$
and $|0,4\rangle$. In this case the superexchange approach involves
$70$ terms with varying signs and
gives $\Delta_{4}\approx 1.4 \times 10^{-5}$ cm$^{-1}$
and the phase space approach
again involves only one term yielding
$\Delta_{4}\approx 2.1 \times 10^{-5}$ cm$^{-1}$. 
For comparison the exact splitting $\approx 1.2 \times 10^{-5}$
cm$^{-1}$.}
\end{thebibliography}
\end{document}